\documentclass[preprint,aps,amsmath,showpacs,floatfix]{revtex4}
\usepackage{graphicx}
\begin{document}

\title{On nonextensive thermo-statistics: systematization,
clarification of scope and interpretation, and illustrations}
\author{Roberto Luzzi, \'Aurea R. Vasconcellos, and J. Galv\~ao Ramos}
\affiliation{Instituto de F\'{\i}sica ``Gleb Wataghin'',
Universidade Estadual de Campinas, Unicamp, 13083-970, Campinas, S\~ao Paulo, Brazil}

\begin{abstract}
When dealing with certain kind of complex phenomena the
theoretician may face some difficulties -- typically a failure to
have access to information for properly characterize the system --
for applying the full power of the standard approach to the well
established, physically and logically sound, Boltzmann-Gibbs
statistics. To circumvent such difficulties, in order to make
predictions on properties of the system and looking for an
understanding of the physics involved (for example in analyzing
the technological characteristics of fractal-structured devices)
can be introduced large families of auxiliary statistics. We
present here a systematization of these different styles in what
can be termed as Unconventional Statistical Mechanics, accompanied
of an analysis of the construction and a clarification of its scope
and interpretation. As illustrations are derived heterotypical
Bose-Einstein, Fermi-Dirac and Maxwell-Boltzmann distributions,
and several applications including studies of experimental works
are briefly described.
\end{abstract}

\maketitle

\section{Introduction}

Over twenty years ago Montroll and Shlesinger wrote that {\it in
the world of investigation of complex phenomena that requires
statistical modeling and interpretation, several competing styles
have been emerging}, each with its own champions \cite{montroll}.
Lately a large amount of efforts -- with a flood of papers related
to one particular case -- has been given to the topic. What is at
play consists in that in the study of certain physicochemical
systems we may face difficulties when handling situations
involving fractal-like structures, correlations (spatial and
temporal) with some type of scaling, turbulent and chaotic motion,
small size (nanometer scale) system with eventually a low number
of degrees of freedom and complicate boundary conditions,
generalized hydrodynamics, and so on. The interest on them has
been recently enhanced as a consequence that such situations are
present in electronic and opto-electronic devices of the nowadays
point-first technologies and also in technological areas involving
the use of disordered systems, polymeric solutions, conducting
glasses, the case of microbatteries, and others. The question is
that these difficulties consist, as a rule, in that the
theoretician cannot properly satisfy {\it Fisher's criteria of
efficiency and sufficiency} \cite{fisher} in the conventional,
well established, physically and logically sound Boltzmann-Gibbs
statistics, meaning an impairment to correctly include the
presence of large fluctuations (and eventually higher-order
variances) and to account for the relevant and proper
characteristics of the system, respectively. Then, just out of
necessity and convenience and to be able to make predictions --
and in the way providing an understanding, even partial, of the
physics of the system of interest, for example in analyzing
technological characteristic of physico-chemical systems as
illustrated in Sect. III -- one may resort to the use of
statistics other than the usual form of the Boltzmann-Gibbs (BG)
one, which are not at all extensions of the latter but, as said,
introduce an auxiliary detour for analyzing the problem in hands.

Among existing approaches we can mention what can be called {\it
Generalized Statistical Mechanics} as used by P. T. Landsberg
showing that functional properties of the (informational)
entropies (see next section) give origin to different types of
thermostatistics, and rise the question of how to select a
``proper'' one, that is, are some better than others?
\cite{landesberg}; for decades has been in use {\it Levy
Statistics}, introducing a modified non-Gaussian distribution
which has been applied to a variety of problems (see for example
Ref. \cite{shlesinger}); the approach of W. Ebeling who has
addressed the question of the treatment of a class of systems, in
nature and society, which are determined by their total history;
it can be referred-to as {\it Ebeling Statistics} \cite{ebeling};
{\it Superstatistics} developed by C. Beck and E. G. D. Cohen for
nonequilibrium systems with complex dynamics in stationary states
with large fluctuations on long-time scales \cite{beck1}; {\it
Non-Extensive Statistics} based on Havrda-Charvat statistics
\cite{havrda} applied to a number of cases are described in the
Conference Proceedings of Ref. \cite{abe}: It consists in the
so-called Tsallis statistics \cite{tsallis}; {\it Renyi
statistics} -- devised by the renowned theoretical statistitian A.
Renyi \cite{renyi} -- has been introduced in the scientific
literature, as noticed in Ref. \cite{takens}, with, for example,
P. Grassberger and I. Procaccia \cite{grassberger} using it as a
valuable method for characterizing experimental chaotic signals,
and recently P. Jizba and T. Arimitzu \cite{jizba} have presented
an extensive analysis of it in a paper called ``The world
according to Renyi''; {\it Sharma-Mittal Statistics} \cite{sharma}
or better to say a variation of it (called {\it Kappa or
Deformational statistics}) was used by V. M. Vasyliunas in
problems of astrophysical plasma \cite{vasylicenas} and by G.
Kaniadakis in the case of relativistic systems \cite{kaniadakis};
and so on.

It has been stated that some of these statistics are {\it
nonextensive} as compared with BG statistics (e.g. Ref.
\cite{abe}) but this is not quite correct once in the strict
equilibrium of adiabatically closed systems, described by the
microcanonical ensemble, the BG entropy is nonextensive only
becoming practically extensive in the {\it thermodynamic limit}
(e.g. Ref. \cite{pathriabook}). The latter is one case of
idealization in science \cite{sklar1}, useful for particular
situations however not applicable in general: notice for example
the case of small systems \cite{hill} present, e.g., in
nano-science and technology which are of a quite large interest
and in ample development nowadays.

We recall that in Statistical Mechanics the probability
distribution (statistical operator), usually derived from
heuristic arguments, can also be derived from a variational
method, once is made connection with Information Theory
\cite{jaynespapers,jaynesbook,luzzi1}. It consists into making
extremal -- a maximum --, subject to certain constraints, a
functional (superoperator) of the probability distribution
(statistical operator). Such quantity, first introduced in
Shannon's Theory of Communication \cite{weaver}, can be referred
to as {\it measure of uncertainty of information}. It has also
been called statistical measure and entropy, with the
understanding that it is {\it informational entropy}. It is worth
to emphasize -- in view of some confusion that has recently
pervaded the scientific literature -- that the different possible
informational entropies are not to be interpreted as the
thermodynamic entropy of the physical system. R. T. Cox has
noticed that the meaning of such entropies is not the same in all
respects as that of anything which has a familiar name in common
use, and it is therefore impossible to give a simple verbal
description of it, which is, at the same time, an accurate
definition \cite{cox}. E. T. Jaynes has also commented that it is
an unfortunate terminology, and a major occupational disease in
that there exists a persistent failure to distinguished between
the informational entropy, which is a property of any probability
distribution, and the experimental entropy of thermodynamics,
which is instead a property of the thermodynamic state: Many
research papers are flawed fatally by the authors' failure to
distinguish between these entirely different things, and in
consequence providing nonsense results \cite{jaynesbook2}. Along
such erroneous line, recently it has been considered the utterly
wrong proposition that a particular informational entropy -- among
the infinitely-many that can be defined -- comes to supersede the
Boltzmann-Gibbs one as the entropy of physical systems
\cite{abe,cho}: Such ``entropy'' has a form of the so-called
structural Havrda-Charvat one \cite{havrda}, which is, we restate,
a generating functional for deriving so-called heterotypical
probability distributions \cite{kapur} to contour, as noticed, the
difficulties of application of Boltzmann-Gibbs statistics when we
face limitation (ours, not of BG statistics) in satisfying
Fisher's criteria in statistics \cite{luzzi2}.

As noticed in the Abstract we present a systematization of these
alternative statistics, dubbed {\it Unconventional Statistical
Mechanics} (USM), within the variational formalism MaxEnt-NESEF
(for Maximization of Informational Entropy in the Non-Equilibrium
Statistical Ensemble Formalism \cite{luzzi1,luzzi3}) with the use
of non-conventional statistical measures (informational entropies)
to derive heterotypical nonequilibrium probability distributions.
This is discussed in the next Section, and after that we
illustrate the matter with the derivation of heterotypical
Fermi-Dirac, Bose-Einstein, and Maxwell-Boltzmann single-particle
distribution functions, and with a brief description of the
application of USM in several situations including analysis of
experimental data (see also Refs. \cite{luzzi4,aurea6247,luzzi5}).

\section{Unconventional Statistical Mechanics}

In Statistical Mechanics the variational approach MaxEnt-NESEF --
mentioned in the Introduction --
\cite{jaynesbook,luzzi1,luzzi3,zubarao} provides a powerful,
practical and soundly based procedure, of a quite broad scope, for
building a nonequilibrium ensemble formalism. It is encompassed in
what is sometimes referred-to as {\it Informational-Entropy
Optimization Principles} (see for example Ref. \cite{kapur}), or,
to be more precise, we would say {\it constrained optimization},
that is, restricted by constraints consisting in the available
information. Such optimization is performed through calculus of
variation with Lagrange's method for finding the constrained
extremum being the preferred one.

In the conventional approach, let it be in the cases of
equilibrium \cite{jaynesprd,elsasser}, or nonequilibrium (in the
linear or Onsagerian regime \cite{grandy}, or for arbitrarily
far-from-equilibrium systems \cite{luzzi1,luzzi3,zubarao}), one
proceeds to maximize Shannon-Jaynes measure of information -- or
informational entropy which in equilibrium is Gibbs entropy giving
rise to Boltzmann entropy in the microcanonical ensemble in
equilibrium --, namely

\begin{equation}
S_{SJ}(t)=-Tr\left\{ \rho (t) \,ln \,\rho(t) \right\},  \label{eq1}
\end{equation}

\noindent under the given constraints. The calculus leads to
statistical operators of an exponential form, with the exponent
depending on the set of basic dynamical variables present in the
constraints ($\{ \hat{P}_j (\mathbf{r})\}$, see below), and the
set of Lagrange multipliers ($\{ F_j (\mathbf{r},t) \}$, see
below), the latter being the nonequilibrium (or equilibrium
according to the case) thermodynamic variables. In fact we must
proceed to the maximization of the SJ-informational entropy with
the constraints

\begin{eqnarray}
Tr\left\{ \rho (t) \right\} =1 \, , \label{eq2}\\
Q_j (\mathbf{r},t) = Tr\left\{ \hat{P}_j (\mathbf{r}) \rho(t)
\right\}, \label{eq3}
\end{eqnarray}

\noindent with $j=1,2,\ldots$. Equation (\ref{eq2}) is the
normalization condition and Eq. (\ref{eq3}) consists in the set of
average values over the nonequilibrium ensemble (characterized by
the sought after statistical operator $\rho(t)$) of the set of
microdynamical variables $\{ \hat{P}_j (\mathbf{r}) \}$ chosen for
providing the constraints in the variational method. The resulting
statistical operator is given by
\cite{luzzi1,luzzi3,zubarao,grandy}

\begin{equation}
\rho_{\epsilon} (t) = \exp \left\{ - \hat{S}(t,0) +
\int_{-\infty}^{t} dt' e^{\epsilon(t'-t)} \frac{d}{dt'} \hat{S}
(t',t'-t) \right\}, \label{eq4}
\end{equation}

\noindent where

\begin{equation}
\hat{S} (t',t'-t)=  \exp{ \left\{ - \frac{1}{i \hbar}
(t'-t)\hat{H} \right\}} \hat{S} (t',0) \exp{ \left\{  \frac{1}{i
\hbar} (t'-t)\hat{H} \right\}}, \label{eq5}
\end{equation}

\noindent and

\begin{equation}
\hat{S} (t,0)=  \phi(t) + \sum_j  \int d^3 r F_j (\mathbf{r},t)
\hat{P}_j (\mathbf{r})  \equiv - ln \, \bar{\rho} (t,0)
\label{eq6}
\end{equation}

\noindent is the so-called informational-statistical-entropy
operator \cite{hassan}, and $\bar{\rho} (t,0)$ is an auxiliary
(sometime called ``instantaneously frozen quasiequilibrium'')
statistical operator having an exponential form resembling a
canonical-like distribution \cite{luzzi3}. In these expressions,
$\hat{H}$ is the system Hamiltonian, $\{ F_j (\mathbf{r},t) \}$
the set of Lagrange multipliers associated to the set of basic
microdynamical variables $\{ \hat{P}_j (\mathbf{r}) \}$, and
$\phi(t)$ ensures the normalization condition of Eq. (\ref{eq2})
and can be considered as being the logarithm of a nonequilibrium
partition function, say, $\phi (t) \equiv ln \, \bar{Z}(t)$.
Moreover, the term containing the positive infinitesimal
$\epsilon$ -- which goes to zero after the calculation of averages
has been performed -- results from introducing Abel's kernel (in
the convergence of integral transforms) in the integral on time
\cite{luzzi1,luzzi3,zubarao}. We recall that it introduces in the
theory the concept of Bogoliubov's quasiaverages \cite{bogoliubov}
leading to irreversible evolution from an initial condition, what
it does by selecting the retarded solutions of the Liouville
equation that $\rho (t)$ satisfies: the advanced solutions are
discarded in a quite similar way as done by Gell-Mann and
Goldberger in the case of Schr\"odinger equation in scattering
theory \cite{gellmann}. The statistical operator of Eq.
(\ref{eq4}) satisfies, as it should, the question of historicity
-- in Kirkwood's and Mori's sense \cite{kirk,mori} -- and
irreversibility is introduced here in the so-called
interventionistic picture in logic, resorting to Krylov's
``jolting'' approach \cite{krylov,sklar2}, using a Poissonian
distribution (Abel's kernel) to account for the nonisolation of
any physical system in the real world
\cite{luzzi1,luzzi3,zubarao}.

In that way it is built a general ensemble formalism in {\it
Boltzmann-Gibbs Statistical Mechanics} for any kind of
thermodynamic state of the system, and fundamental for the study
of the physics of condensed matter: it allows to describe
successfully any possible physical situation in many-body systems.
Of course, such success is possible if we are able to properly
handle it, what means, as already mentioned, that in any problem
we need to satisfy Fisher's criteria \cite{fisher} for it to give
satisfactory predictions. As noticed, we may face difficulties for
some kind of situations where the use of BG statistics is simply
impaired because of either becomes not possible to handle the
required {\it information relevant} to the problem in hands, or,
more important and fundamental, is involved a failure on the part
of the theoretician to have a correct access to such information.
Typical cases are that of small systems (few particles) when we
may not be satisfying Fisher's criterion of efficiency
(fluctuations and higher-order variances cannot be ignored), and
the case of complex systems with some type or other of
fractal-like structures or with long-range space correlations or
particular long-time correlations, when we have deficiencies in
the proper knowledge of the characterization of the states of the
system (failure to satisfy Fisher's criterion of sufficiency). We
reemphasize that what is present is a practical difficulty (a
limitation on the part of the theoretician) in the otherwise
complete, physically and logically sound BG-statistics, and this
is illustrated in Section III.

When facing such difficulties, out of necessity (obstacles in
accounting for the relevant information) or, mainly, the force of
circumstances (lack of the proper characterization of the system)
a way to circumvent them has consisted into resorting to {\it
auxiliary (or unconventional) statistics} (see the first sentence
in the Introduction).

This auxiliary Unconventional Statistical Mechanics consists of
two steps:

\begin{enumerate}
\item The choice of a {\it heterotypical probability
distribution}, obtained through application of MaxEnt to different
measures of information, that is, other than the Shannon-Jaynes
one that leads to the conventional BG-statistics; \item The use of
the {\it escort probability} in terms of the chosen heterotypical
probability of item 1 above, in the calculation of averages.
\end{enumerate}

Let us consider both concepts. The escort probability  of order
$\gamma$ of a given probability distribution $\rho$ is defined as

\begin{equation}
\mathcal{D}_{\gamma} \{ \rho \} = \rho^{\gamma} / Tr \left\{
\rho^{\gamma}\right\} , \label{eq7}
\end{equation}

\noindent where $\gamma$ is a positive real number, which seems to
have been introduced by A. Renyi but in terms of Renyi's
heterotypical probability distribution (see Ch. 10 in the book of
Ref. \cite{renyi}), being generalized, and the name given, by C.
Beck and F. Schl\"ogl (see Ch. 5 in the book of Ref.
\cite{beck2}). Its role is to add to the normal definition of
average value the presence of second and higher-order variances
(in that way trying to improve upon the possible failure of
distribution $\rho$ to satisfy Fisher's criterion of efficiency):
For the average value of an observable $\hat{A}$, if we write
$\gamma = 1 + \epsilon$ and make an expansion in powers of
$\epsilon$ we find that

\begin{eqnarray}
\left< \hat{A} \right>_\gamma &=& Tr \left\{ \hat{A} \mathcal{D}_{\gamma}
\left\{ \rho \right\} \right\} \nonumber \\
&=& Tr \left\{ \hat{A} \rho \right\} + \epsilon \left[ \left< \hat{A} \hat{S}
\right> - \left< \hat{A}  \right>\left<  \hat{S} \right> \right] \nonumber \\
&+& \frac{\epsilon^2}{2} \left[ \left< \hat{A} \hat{S}
\hat{S}\right> - \left< \hat{A}  \right>\left<  \hat{S} \hat{S}
\right> + 2 \left< \hat{A}  \right>\left<  \hat{S}  \right>^2 - 2
\left< \hat{A} \hat{S} \right>\left<  \hat{S}  \right>  \right] +
\mathcal{O} (\epsilon^3), \label{eq8}
\end{eqnarray}

\noindent where

\begin{equation}
\left< \ldots \right> = Tr \left\{ \ldots \rho \right\}, \label{eq9}
\end{equation}

\noindent stands for the normal average. For illustration let us
take for $\hat{A}$ the Hamiltonian $\hat{H}$ and for $\rho$ a
canonical distribution $\rho = Z^{-1} \exp{ \{- \beta \hat{H} \}
}$, and then using Eq. (\ref{eq8}) up to second order in $\epsilon
= \gamma -1$ it follows that the energy is given by

\begin{equation}
E= \left< \hat{H}\right>_\gamma = \left< \hat{H}\right> + \epsilon
\beta \Delta_2 E + \frac{\epsilon^2}{2} \beta^2 \Delta_3 E,
\label{eq10}
\end{equation}

\noindent where

\begin{eqnarray}
\Delta_2 E &=& \left< \left( \hat{H} - \left< \hat{H}
\right>\right)^2\right> = \left< \hat{H}^2 \right> - \left<
\hat{H} \right>^2, \label{eq11}\\
\Delta_3 E &=& \left< \left( \hat{H} - \left< \hat{H}
\right>\right)^3\right> = \left< \hat{H}^3 \right> - 3\left<
\hat{H} \right>\left< \hat{H}^2 \right> + 2 \left< \hat{H}
\right>^3, \label{eq12}
\end{eqnarray}

\noindent with $\Delta_2 E$ and $\Delta_3 E$ being the second and
the third order variances of the energy. The departure of $\gamma$
from the value 1 gives an indication of the influence of the
variances in the process of prediction of average values.

On the other hand, for obtaining the heterotypical distributions
one needs to introduce alternative measures of information (i.e.
informational entropies). A large family is the one provided by I.
Csiszer (\cite{csiszer} and see also the book in Ref.
\cite{kapur}) and, among the infinitely-many possibilities, we
list in Table I five of them ($W^{-1}$ is the constant probability
in the uniform distribution). K\"ulback-Leibler measure
\cite{kulback}, which is parameter independent, corresponds to
Shannon-Jaynes informational entropy and produces the usual
results in BG-statistics. The others are dependent on what has
been dubbed as {\it infoentropic index(es)}: The one of the
Havrda-Charvat \cite{havrda} has been used in Physics and other
disciplines with the name of Tsallis entropy \cite{tsallis}
(calling $q$ the infoentropic index); Sharma-Mittal informational
entropy \cite{sharma} depends on two infoentropic indexes  (taking
index $\beta=1$ it goes over the one of Havrda-Charvat, and for a
particular relation between both indexes results Kappa-statistics
of Refs. \cite{vasylicenas,kaniadakis}), which is a kind of
weighted superposition of two Havrda-Charvat measures. Renyi
measure \cite{renyi} has been used in a number of problems in
several disciplines, including Physics
\cite{takens,jizba,grassberger} (it can be noticed that Renyi and
Havrda-Charvat measures produce, via MaxEnt, the same statistical
operator, differing only in the interpretation of the Lagrange
multipliers). Kapur measure \cite{kapur2}, depending on two
infoentropic indexes is a superposition of two Renyi measures.

Two important points need be stressed: On the one hand,
Shannon-K\"ulback-Leibler measure is derived from a set of
fundamental axioms, and Havrda-Charvat and Renyi measures follow
from the modification of one of such axioms
\cite{kulback,havrda,renyi}. On the other hand, in the process is
introduced an open parameter (the infoentropic index $\alpha$,
recently indicated by $q$), to be determined by fitting of the
index-dependent prediction with observation.

The role of these heterotypical probability distributions is to
introduce a selective weighting of, for example, occupation
functions of quantum states, amplitudes of movement in hydrodynamics,
etc. This is illustrated in next Section.\\ \\

\begin{tabular}{|l|c|}
\multicolumn{2}{c}{Table I: Informational-Statistical Entropies} \\
\hline
\hline
\multicolumn{2}{|l|}{Conventional ISE} \\
\hline
\hline
Boltzmann-Gibbs-Shannon-Jaynes ISE   &     $-Tr \{ \rho \, ln \, \rho \}$ \\
(from K\"ulback-Leibler measure)     &                              \\
\hline
\hline
\multicolumn{2}{|l|}{Unconventional (entropic-indexes-dependent) ISEs} \\
\hline
\hline
From Havrda-Charvat measure   & $-\frac{1}{\alpha -1} Tr \{ \rho^{\alpha} - \rho \}$ \\
                              & $\alpha > 0$ and $\alpha \neq 1$ \\
\hline
From Sharma-Mittal measure    & $-\frac{W^{\beta -1}}{\alpha -\beta}
Tr \{ [W^{\alpha-\beta}\rho^{\alpha-\beta+1} - \rho] \rho^{\beta-1}\}$ \\
                              & $\alpha > 1$, $\beta \le 1$ or $\alpha < 1$, $\beta \ge 1$ \\
\hline
From Renyi measure            & $-\frac{1}{\alpha -1} ln \, Tr \{ \rho^{\alpha} \}$ \\
                              & $\alpha > 0$ and $\alpha \neq 1$\\
\hline
From Kapur Measure            & $-\frac{1}{\alpha -\beta} [ ln \,
Tr \{ \rho^{\alpha}\} - ln \, Tr \{ \rho^{\beta}\}]$ \\
                              & $\alpha > 0$, $\beta >0$ and $\alpha \neq \beta$\\
\hline
\end{tabular}

\section{Illustrative Examples}

In this Section we illustrate the use of the Unconventional
Statistical Mechanics described in the previous ones. We present
comments on a few cases and is given the reference where details
can be consulted. One example is the case of ``anomalous''
luminescence in semiconductor heterostructures, which requires the
introduction of heterotypical Fermi-Dirac distributions. We begin
with this case.

\subsection{Photoluminescence in Semiconductor Heterostructures}

There exists nowadays a large interest on the question of optical
properties of quantum wells in semiconductor heterostructures,
which have been extensively investigated in the last decades as
they have large relevance for the high performance of electronic
and optoelectronic devices (see for example Ref. \cite{singh}). To
deal with these kind of systems, because of the constrained
geometry that they present (where phenomena develop in nanometer
scales) the researcher has to face difficulties with the
theoretical analysis: A most relevant question to be dealt with is
the one related to the interface roughness, usually having a kind
of fractal-like structure which leads to energies and wavefunction
depending on boundary conditions which need account for spatial
correlations. As a consequence the different physical properties
of these systems appear as, say, ``anomalous'' when the results
are compared with those that are observed in bulk materials,
particularly, the case of photoluminescence which we briefly
describe here. In the study of photoluminescence in nanometric
quantum wells the conventional treatment via the well established
Boltzmann-Gibbs formalism has its application impaired because of
the spatial correlations resulting from the non smooth
confinement, relevant in the characterization of the system, on
which one does not have access to (obviously the interface
roughness varies from sample to sample and one does not have any
easy possibility to determine the topography of the interface).
This is then the reason why the {\it criterion of sufficiency} is
not satisfied in this case.

Let us consider a system of carriers (electrons and holes)
produced, in the quantum well of an heterostructure, by a
ultrafast laser pulse. They are out of equilibrium and their
nonequilibrium macroscopic state can be described in terms of a
nonequilibrium statistical thermodynamics \cite{luzzi1}, with the
nonequilibrium thermodynamic state characterized by the time
evolving quantities energy and density (see for example Ref.
\cite{algarte}). Electrons and holes do recombine producing a
luminescence spectrum which, we recall, is theoretically expressed
as

\begin{equation}
I(\omega |t) \propto \sum_{n,n',\mathbf{k}_{\perp}}
f^{e}_{n\mathbf{k}_{\perp}} (t) \, f^{h}_{n'\mathbf{k}_{\perp}}
(t) \delta \, (\hbar \Omega - \epsilon^{e}_{n\mathbf{k}_{\perp}} -
\epsilon^{h}_{n\mathbf{k}_{\perp}}), \label{eq13}
\end{equation}

\noindent where $f^e$ and $f^h$ are the populations of electrons
and holes, $\hbar \Omega = \hbar \omega - E_G$, with $\omega$
being the frequency of the emitted photon and $E_G$ the energy
gap, and $\epsilon^{e(h)}_{n\mathbf{k}_{\perp}}$ are the electron
(hole) individual energy levels in the quantum well (index $n$ for
the discrete levels and $\mathbf{k}_{\perp}$ for the free movement
in the $x - y$ plane), and the delta function accounts for energy
conservation.

The textbook expression for the energy levels corresponding to the
use of perfectly smooth bidimentional boundaries is given by

\begin{equation}
\epsilon^{e(h)}_{n\mathbf{k}_{\perp}}= n^2 \frac{\pi^2 \hbar^2
}{2m^*_{e(h)} L^2_{QW}} + \frac{\hbar^2 \mathbf{k}^2_{\perp}
}{2m^*_{e(h)} } \, , \label{eq14}
\end{equation}

\noindent where $L_{QW}$ is the quantum-well width and
$m^*_{e(h)}$ is the effective mass and the populations $f$ take a
form that resembles instantaneous in time Fermi-Dirac
distributions: see Ch. 6 in the book of Ref. \cite{luzzi1} and
Ref. \cite{algarte}. Using this expression in the calculation of
$I$ of Eq. (\ref{eq13}) and comparing it with the experimental
results \cite{laureto,aurea;laureto} one finds a disagreement, and
then it is used the name of ``anomalous'' luminescence for these
experimental results. This is a consequence that we are using an
improper description of the carriers' energy levels -- we are not
satisfying the {\it criterion of sufficiency} (as discussed in
previous Sections) --, resulting of ignoring the roughness of the
boundaries (with self-affine fractal structure \cite{family})
which needs be taken into account in these nanometric-scale
geometries, and then the boundary conditions to be placed on the
wavefunctions are space dependent. Hence complicated space
correlations are to be introduced, but to which we do not have
access (information), as already noticed. Hence, this limitation
on the part of the researcher breaks the sufficiency criterion,
and application of the Boltzmann-Gibbs-Shannon-Jaynes construction
is impaired, and one can try to circumvent the difficulty
introducing unconventional statistics based on parameter-dependent
structural informational entropies in order to analyze the
properties of the system.

\subsubsection{Distribution Fuctions of Fermions and Boson in USM}

According to Eq. (\ref{eq13}) what we need is to express the
carrier populations in a heterotypical statistics, and we resort
to the Renyi one (see for example Ref. \cite{jizba} and
\cite{luzzi4,aurea6247,luzzi5}). Using MaxEnt-NESEF in terms of
Renyi statistical entropy it can be derived the corresponding
auxiliary (``instantaneously frozen'') statistical operator
$\bar{\rho} (t,0)$ [cf. Eq. (\ref{eq6}) -- the case of the
conventional one --, and we recall that the statistical operator
is given in term of this auxiliary one in Eq. (\ref{eq4})]. After
some straightforward mathematical manipulations it follows that it
can be written in a convenient form for performing calculations,
namely, \cite{aurea6247}

\begin{equation}
\bar{\rho}_\alpha (t,0) = \frac{1}{\tilde{\eta}_\alpha (t)} \left[
1 + (\alpha -1) \sum_j \tilde{F}_{j \alpha} (t)
\hat{P}_j\right]^{-\frac{1}{\alpha -1}} , \label{eq15}
\end{equation}

\noindent where

\begin{eqnarray}
\tilde{\eta}_\alpha (t) = Tr \left\{ \left[ 1 + (\alpha -1) \sum_j
\tilde{F}_{j \alpha} (t) \hat{P}_j\right]^{-\frac{1}{\alpha -1}}
\right\}, \label{eq16}\\
\tilde{F}_{j \alpha} (t) = {F}_{j \alpha} (t) \left[ 1 - (\alpha
-1) \sum_m {F}_{m \alpha} (t) Q_m (t) \right]^{-1}.  \label{eq17}
\end{eqnarray}

Equation (\ref{eq16}) stands for a modified form of the quantity
that ensures the normalization condition, and Eq. (\ref{eq17}) for
modified Lagrange multipliers, where $F_{m \alpha}$ are the
original Lagrange multipliers, and $Q_m$ the average values of the
microdynamical variables $\hat{P}_m$. Moreover, we recall that in
Renyi statistics the associated escort probability is given by

\begin{equation}
\mathcal{D}_\alpha \{ \rho_\alpha \} = \rho_{\alpha}^{\alpha} / Tr
\{ \rho^{\alpha}_{\alpha} \}, \label{eq18}
\end{equation}

\noindent that is, the order of the escort probability is the same
as the index in Renyi heterotypical distribution (see Ch. 10 in
the book of Ref. \cite{renyi}). In particular to derive the {\it
distribution functions for fermions and for bosons} using USM in
terms of Renyi statistical approach one chooses as basic dynamical
variables, i.e. the $\hat{P}_j$ in Eq. (\ref{eq15}), the set of
occupation number operators

\begin{equation}
\{ \hat{n}_{\mathbf{k}} \} = \{ c^{\dag}_{\mathbf{k}}
c_{\mathbf{k}}  \} , \label{eq19}
\end{equation}

\noindent where $c \, (c^{\dag})$ are the usual annihilation
(creation) operators in states $| \mathbf{k} \rangle$, satisfying
the corresponding commutation and anticommutation rules of,
respectively, bosons and fermions (the spin index is ignored).
Their average values are the infoentropic-index $\alpha$-dependent
distribution functions

\begin{equation}
f_{\mathbf{k}}(t) = Tr \left\{ c^{\dag}_{\mathbf{k}}
c_{\mathbf{k}} {\mathcal{D}}_{\alpha \epsilon}\{  {\rho}_{\alpha
\epsilon}\ (t) \} \right\}= Tr \left\{ c^{\dag}_{\mathbf{k}}
c_{\mathbf{k}} \bar{\mathcal{D}}_\alpha \{  \bar{\rho}_\alpha
(t,0) \} \right\} , \label{eq20}
\end{equation}

\noindent which is a consequence that for the basic variables, and
only for the basic variables, the average value coincides with the
one calculated with the auxiliary distribution
\cite{luzzi1,zubarao}. The auxiliary statistical operator is then
[cf. Eq. (\ref{eq15})]

\begin{equation}
\bar{\rho}_\alpha (t,0) = \frac{1}{\tilde{\eta}_\alpha (t)} \left[
1 - (\alpha -1) \sum_{\mathbf{k}} \tilde{F}_{\mathbf{k} \alpha}
(t) c^{\dag}_{\mathbf{k}} c_{\mathbf{k}} \right]^{-\frac{1}{\alpha
-1}} , \label{eq21}
\end{equation}

\noindent with [cf. Eq. (\ref{eq17})]

\begin{equation}
\tilde{F}_{\mathbf{k} \alpha} (t) = {F}_{\mathbf{k} \alpha} (t)
\left[ 1 + (\alpha -1) \sum_{\mathbf{k}'} {F}_{ \mathbf{k}'
\alpha} (t) f_{\mathbf{k}'} (t) \right]^{-1}.  \label{eq22}
\end{equation}

The populations of Eq. (\ref{eq20}), according to the calculation
described in Ref. \cite{luzzi4} take the form

\begin{equation}
f_{\mathbf{k}} (t) = \bar{f_{\mathbf{k}}} (t) +
\mathcal{C}_{\mathbf{k}} (t) , \label{eq23}
\end{equation}

\noindent where

\begin{equation}
\bar{f}_{\mathbf{k}} = \frac{1}{ \left[ 1 + (\alpha -1)
\tilde{F}_{\mathbf{k} \alpha} (t)
\right]^{\frac{\alpha}{\alpha-1}} \pm 1  } , \label{eq24}
\end{equation}

\noindent where upper plus sign stands for fermions, and the lower
minus sign for bosons, and

\begin{equation}
\mathcal{C}_{\mathbf{k}} (t) = \alpha (1 - \alpha) (1 -
\bar{f}_{\mathbf{k}}(t)) \sum_{\mathbf{k}} \tilde{F}_{\mathbf{k}
\alpha} (t) \tilde{F}_{\mathbf{k}' \alpha} (t) Tr \left\{
c^{\dag}_{\mathbf{k}} c_{\mathbf{k}} c^{\dag}_{\mathbf{k}'}
c_{\mathbf{k}'}  \bar{\mathcal{D}}_\alpha \{ \bar{\rho} (t,0) \}
\right\} + \ldots , \label{eq25}
\end{equation}

\noindent involving two, three, etc. particle correlations, which
in general are minor corrections to the first, and main,
contribution, the one given by Eq. (\ref{eq24}).

In the limit of $\alpha$ going to 1, which applies when the
criteria of efficiency and sufficiency are satisfied, Renyi
statistical entropy acquires the form of the
Boltzmann-Gibbs-Shannon-Jaynes one, $\mathcal{C}$ becomes null,
$\tilde{F}_{\mathbf{k} \alpha} (t)$ becomes ${F}_{\mathbf{k} }
(t)$, and then

\begin{equation}
f_{\mathbf{k}} (t) = \frac{1}{ e^{F_{\mathbf{k}} (t)} \pm 1  } .
\label{eq26}
\end{equation}

\noindent (In equilibrium $F_{\mathbf{k}} (t) \rightarrow (
\epsilon_{\mathbf{k}} - \mu)/ k_B T$ and there follows the
traditional Fermi-Dirac and Bose-Einstein distributions).

We can see that the distribution of Eq. (\ref{eq23}) is composed
of a term $\bar{f}$ corresponding to the individual particle in
state $|\mathbf{k} \rangle$, plus the contribution $\mathcal{C}$
containing correlations (of order two, three, etc.) among the
individual particles. This type of calculation but for systems in
equilibrium and not using the average value defined in Eq.
(\ref{eq8}), in terms of the escort probability, was reported in
Ref. \cite{buy}, and then are not satisfactory.

Let us now give some attention to the Lagrange multipliers
$F_{\mathbf{k} \alpha}(t)$. The most general statistical operator
for nonequilibrium systems can be expressed in the form of a
generalized nonequilibrium grand-canonical statistical operator
for a system of individual quasiparticles, where the basic
variables are independent linear combinations of the
single-quasiparticle occupation number operators [cf. Eq.
(\ref{eq19})], consisting of the energy and particle densities and
their fluxes of all order \cite{luzzi1,luzzi3}. This description
follows from the choice \cite{luzzi1}

\begin{eqnarray}
\tilde{F}_{\mathbf{k} \alpha} (t) &=& \tilde{\beta}_\alpha (t) [
\epsilon_{\mathbf{k}} - \tilde{\mu}_\alpha (t) ] -
\tilde{\mathbf{F}}_{h \alpha} (t) \cdot \epsilon_{\mathbf{k}}
\mathbf{u}(\mathbf{k}) - \tilde{\mathbf{F}}_{n \alpha} (t)
\cdot  \mathbf{u}(\mathbf{k})  \nonumber \\
&-& \sum_{r \ge 2} \left[ \tilde{F}^{[r]}_{h \alpha} (t) \otimes
\epsilon_{\mathbf{k}} u^{[r]} (\mathbf{k}) +  \tilde{F}^{[r]}_{n
\alpha} (t) \otimes u^{[r]} (\mathbf{k})\right] , \label{eq27}
\end{eqnarray}

\noindent where has been introduced the scalar quantities
$\tilde{\beta}(t)$ and $\tilde{\mu}_{\alpha} (t)$,
$\tilde{\mathbf{F}}_{h \alpha} (t)$ and $\tilde{\mathbf{F}}_{n
\alpha} (t)$ are vectors, and $\tilde{F}^{[r]}_{h \alpha}$ and
$\tilde{F}^{[r]}_{n \alpha}$ $r$-th rank tensors. Moreover,

\begin{equation}
u^{[r]} (\mathbf{k}) = [\mathbf{u}(\mathbf{k}) \ldots (r \; \;
\mathrm{- \; \; times}) \ldots \mathbf{u}(\mathbf{k})]\,,
\label{eq28}
\end{equation}

\noindent is the tensorial product of $r$-times the characteristic
velocity $\mathbf{u}(\mathbf{k}) = \hbar^{-1} \nabla_{\mathbf{k}}
\epsilon_{\mathbf{k}}$, where $\epsilon_{\mathbf{k}}$ is the
energy dispersion relation of a single-particle, and then
$\mathbf{u}(\mathbf{k})$ is the group velocity in state
$|\mathbf{k}\rangle$. Dot stands as usual for scalar product of
vectors, and $\otimes$ for fully contracted product of tensors.

To better illustrate the matter, we introduce a simplified
description, or better to say a quite truncated description,
proceeding to neglect in Eq. (\ref{eq27}) all the contributions
arising out of the fluxes, i.e. we put $\mathbf{F}=0$ and
$F^{[r]}=0$, retaining only the first term on the right-hand side.
Therefore, we do have that

\begin{equation}
\bar{f}_{\mathbf{k}} (t) = \frac{1}{\left[
1+(\alpha-1)\tilde{\beta}_\alpha (t) [\epsilon_{\mathbf{k}} -
\tilde{\mu}_\alpha (t)]\right]^{\frac{\alpha}{\alpha-1}}  \pm 1}
\, , \label{eq29}
\end{equation}

\noindent where [cf. Eq. (\ref{eq22})]

\begin{equation}
\tilde{\beta}_\alpha (t) = \beta_\alpha (t) / \{ 1-(\alpha-1)
\beta_\alpha (t) [E(t) - \mu_\alpha (t) N (t)]\}\, , \label{eq30}
\end{equation}

\noindent and $\tilde{\beta} \tilde{\mu} = \tilde{\beta} \mu$,
then $\tilde{\mu} = \mu$ having the role of a quasichemical
potential.

In this Eq. (\ref{eq30}) $E(t)$ is the energy

\begin{equation}
E(t) \simeq \sum_{\mathbf{k}} \epsilon_{\mathbf{k}}
\bar{f}_{\mathbf{k}}(t) , \label{eq31}
\end{equation}

\noindent and $N$ the number of particles

\begin{equation}
N(t) \simeq \sum_{\mathbf{k}} \bar{f}_{\mathbf{k}}(t) \, ,
\label{eq32}
\end{equation}

\noindent where the correlations present in $\mathcal{C}$ in Eq.
(\ref{eq23}) have been neglected. Moreover, in many cases we can
use an approximate expression for the populations, that is, in the
one of Eq. (\ref{eq24}) we admit that $\pm 1$ can be neglected in
comparison with the other term. This is considered as taking a
statistical nondegenerate limit, once, if we put $\alpha $ going
to 1 (what, we again stress, strictly corresponds to the situation
when the principle of sufficiency is satisfied), the population
takes the form of a Maxwell-Boltzmann distribution. In this
condition the expression for the population can be written as

\begin{equation}
\bar{f}_{\mathbf{k}} (t) = A_\alpha (t) [1+(\alpha-1)
\mathcal{B}_\alpha (t)
\epsilon_{\mathbf{k}}]^{-\frac{\alpha}{\alpha -1}} \, ,
\label{eq33}
\end{equation}

\noindent where

\begin{equation}
A_\alpha (t) = [1-(\alpha-1) \tilde{\beta}_\alpha (t)
\tilde{\mu}_\alpha (t)]^{-\frac{\alpha}{\alpha -1}} \, ,
\label{eq34}
\end{equation}

\begin{equation}
\mathcal{B}_\alpha (t) =  \tilde{\beta}_\alpha (t)/ [1-(\alpha-1)
\tilde{\beta}_\alpha (t) \tilde{\mu}_\alpha (t)] \, . \label{eq35}
\end{equation}

Consider a parabolic dispersion relation, that is,
$\epsilon_{\mathbf{k}} = \hbar^2 k^2 / 2m^*$. Using Eq.
(\ref{eq33}) in Eqs. (\ref{eq31}) and (\ref{eq32}), we arrive at
the result that

\begin{eqnarray}
n(t)&=&\frac{N(t)}{V} = A^{3/2}_\alpha (t) \frac{\lambda^{-3}_\alpha (t)}
{4\pi^2} I_{1/2} (\alpha) \,, \label{eq36}\\ \nonumber \\
e(t)&=&\frac{E(t)}{V} = n (t) \frac{I_{3/2} (\alpha) }{I_{1/2}
(\alpha) } k_B \mathcal{T}_\alpha (t) \,, \label{eq37}
\end{eqnarray}

\noindent with the integrals $I_\nu (\alpha)$ related to Beta
functions \cite{luzzi4}, and we have introduced the definition

\begin{equation}
\mathcal{B}^{-1}_{\alpha} (t) = k_B \mathcal{T}_\alpha (t) \, ,
\label{eq38}
\end{equation}

\noindent where $\mathcal{T}$ plays the role of a
pseudotemperature and where $\lambda_\alpha$ in Eq. (\ref{eq36})
is a characteristic length given by $\lambda^2_\alpha (t) =
\hbar^2 / m^* k_B \mathcal{T}_\alpha (t)$ (that is, de Broglie
wave length for a particle of mass $m^*$ and energy $k_B
\mathcal{T}_\alpha (t)$).

We can see that the above Eqs. (\ref{eq36}) and (\ref{eq37})
define the Lagrange multipliers, $\tilde{\beta}_\alpha (t)$ and
$\tilde{\mu}_\alpha (t)$ -- present in $A_\alpha (t), \,
\lambda_\alpha (t)$, and $\mathcal{B}_\alpha (t)$ -- in terms of
the basic variables energy and number of particles. Moreover,
using Eq. (\ref{eq36}) we can obtain an expression for the
quasi-chemical potential in terms of quasitemperature and density,
namely

\begin{equation}
1-(\alpha-1)\tilde{\beta}_\alpha (t) \tilde{\mu}_\alpha (t) =
\left[ 4 \pi^2 \lambda^3_\alpha (t) / I_{1/2} (\alpha)
\right]^{\frac{2(\alpha-1)}{\alpha-3}}
[n(t)]^{\frac{2(\alpha-1)}{\alpha-3}} \, . \label{eq39}
\end{equation}

Also, it can be noticed that for $\alpha=1$ (the case when the
condition of sufficiency is satisfied) one recovers the equivalent
of the results of conventional nonequilibrium statistical
mechanics \cite{aurea6247,grandy}, which are

\begin{equation}
e(t)=\frac{3}{2} n(t) k_B T^*(t) \, , \label{eq40}
\end{equation}

\noindent where we have introduced the so-called quasitemperature
\cite{luzzi1,luzziJCP97,casas,luzzi51}, defined by $k_B T^* =
\mathcal{B}^{-1}_{\alpha=1} (t)$, this equation standing for a
kind of equipartition of energy at time $t$, and

\begin{equation}
\mu (t) = -  k_B T^* (t) \, ln [ T^* (t) / \theta_{tr} (t) ] \, ,
\label{eq41}
\end{equation}

\noindent where $\theta_{tr} (t) = \hbar^2 n^{2/3} (t)  / 2m^*$ is
the characteristic temperature (here in nonequilibrium conditions
and at time $t$) for translational motion. This suggests to define
a so-called ``kinetic temperature'' $\Theta_K (t)$
\cite{nettleton} by equating  $e(t)$ to $(3/2) n(t) k_B \Theta_K
(t)$, given, after Eq. (\ref{eq37}) is used, by

\begin{equation}
\Theta_K (t) = \mathcal{T}_\alpha (t)/(5 -3 \alpha) \, , \label{eq42}
\end{equation}

\noindent where we can see that $\alpha$ must be smaller than
$5/3$, as shown in Ref. \cite{aurea6247} where connection of
theory with experiment is presented, together with other
illustrations and discussions.

\begin{figure}
\hspace{-2cm}\includegraphics[angle=-90,width=18cm]{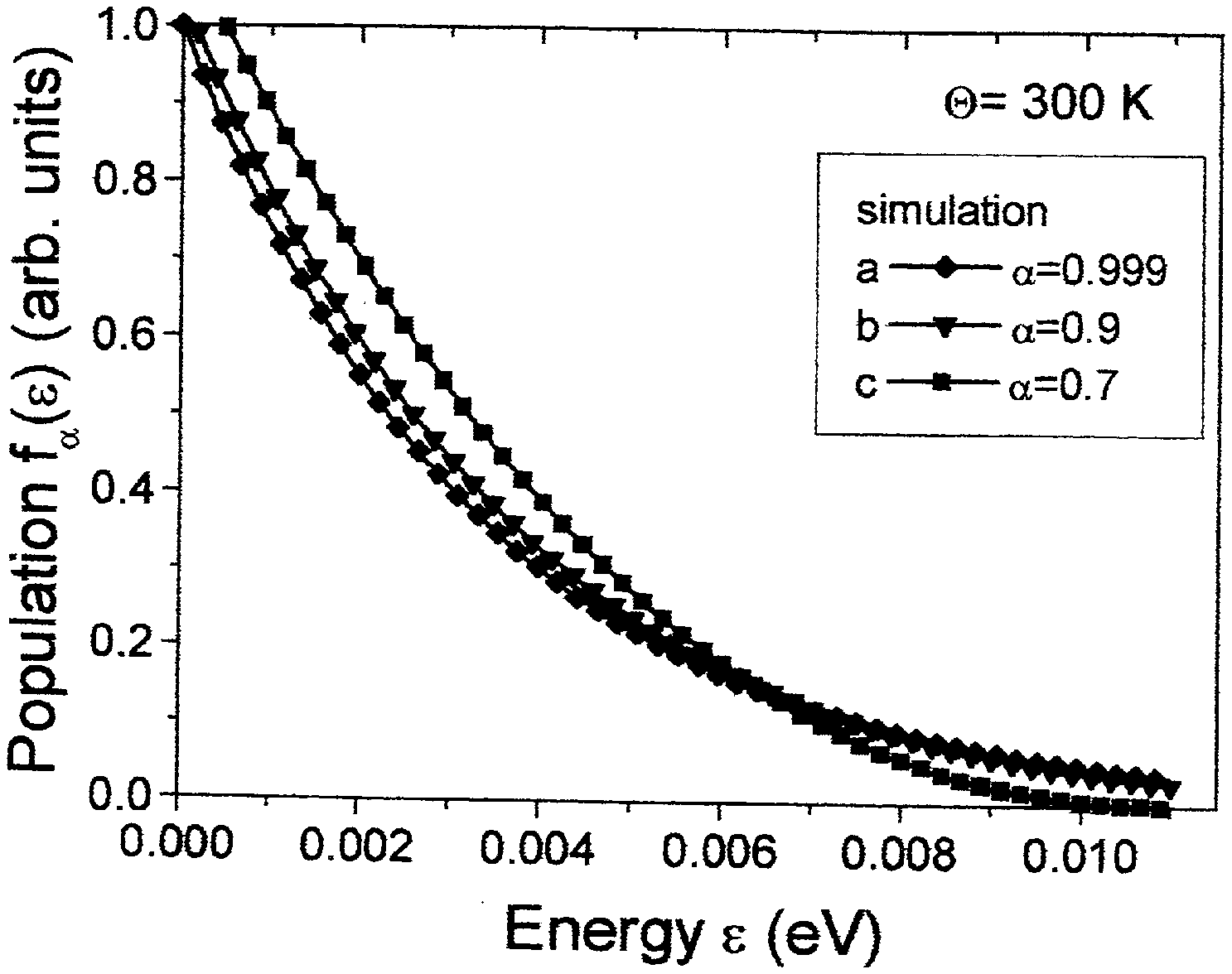}
\caption{The distribution of Eq. (\ref{eq33}) for a
kinetic temperature [cf. Eq. (42)] of $300K$ and values of Renyi's
infoentropic-index $\alpha$ smaller than 1.} \label{fig1}
\end{figure}

\begin{figure}
\hspace{-2cm}\includegraphics[angle=-90,width=18cm]{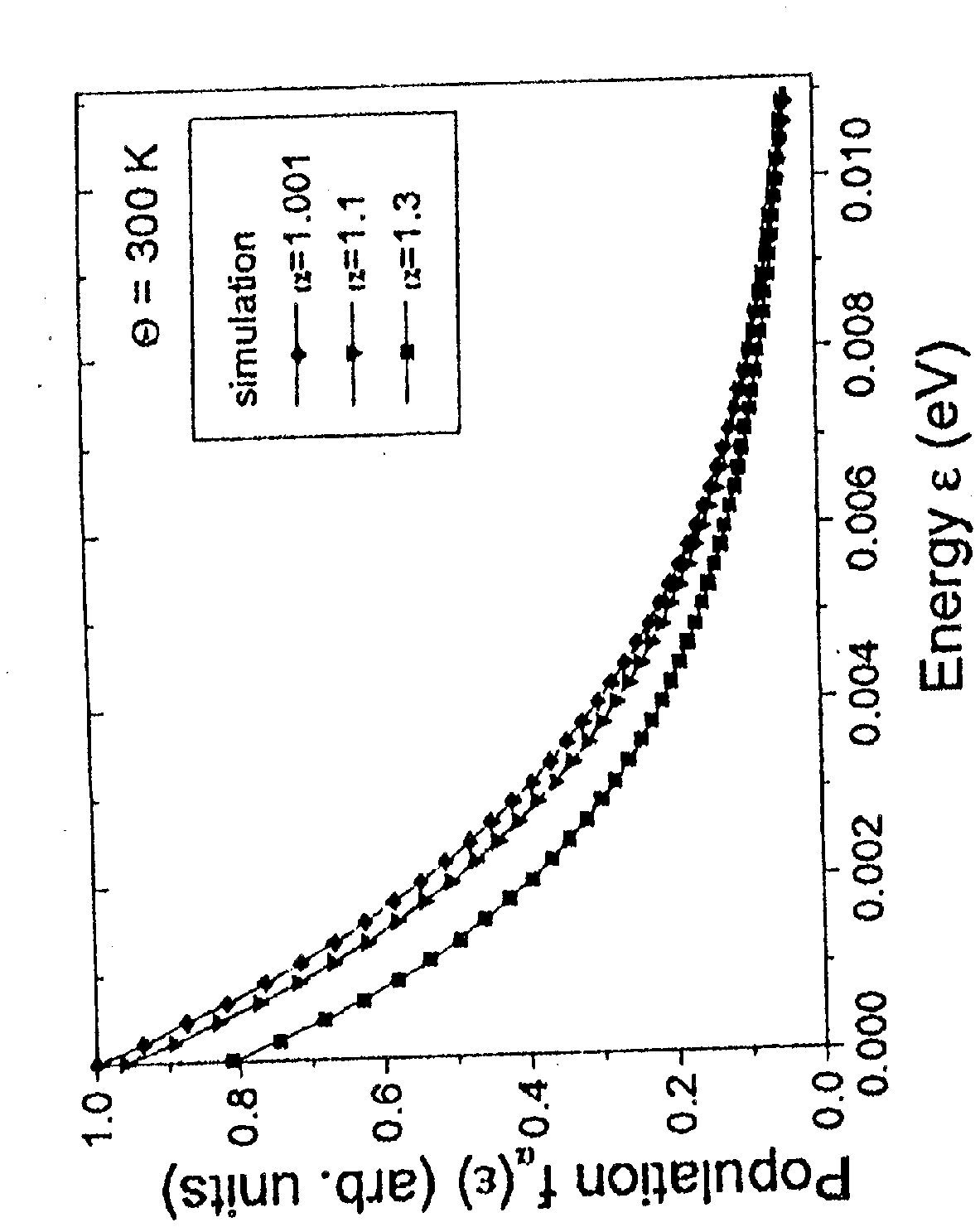}
\caption{The distribution of Eq. (\ref{eq33}) for a kinetic
temperature [cf. Eq. (42)] of $300K$ and values of Renyi's
infoentropic-index $\alpha$ larger than 1.} \label{fig2}
\end{figure}

Let us consider how does the $\alpha$-dependent distribution of
Eq. (\ref{eq29}) compares with the usual Fermi-Dirac and
Bose-Einstein distributions. For illustration we consider the
nondegenerate limit of Eq. (\ref{eq33}), common to both, where
parameter $\mathcal{B}$ is related to the kinetic temperature
$\Theta_K$ by Eqs. (\ref{eq38}) and (\ref{eq42}). Taking for
$\mathcal{T}_\alpha$ of Eq. (\ref{eq38}) the unique value of
$300K$, Figs. \ref{fig1} and \ref{fig2} show the population of Eq.
(\ref{eq33}) corresponding to several values of the infoentropic
index $\alpha$. It can be noticed that the formalism introduces
the characteristic of a different weighting of the values of the
standard distribution $(\alpha \simeq 1)$, such that: (1) for
$\alpha <1$ the population of the modes at low energies are
increased at the expense of those of higher energies $(\epsilon >
7 \times 10^{-3} eV)$, while (2) for $\alpha >1$ we can see the
opposite behavior.

\subsubsection{The "Anomalous" Photoluminescence Spectrum}

Returning to the question of ``anomalous'' luminescence in
nanometric quantum wells in semiconductor heterostructures, the
carrier's populations, to be used in Eq. (\ref{eq13}), are of the
form of Eq. (\ref{eq29}), namely

\begin{equation}
\bar{f}^{e(h)}_{n \mathbf{k}_{\perp}, \alpha} (t) = \left\{
\left\{  1+ (\alpha -1) \tilde{\beta}^{e(h)}_{\alpha} (t) \left[
{\epsilon}^{e(h)}_{n \mathbf{k}_{\perp}} - {\mu}^{e(h)}_{\alpha}
(t) \right]  \right\}^{\frac{\alpha}{\alpha -1}}  \pm 1
\right\}^{-1} , \label{eq43}
\end{equation}

\noindent which depend on $\epsilon^{e(h)}_{n
\mathbf{k}_{\perp}}$, that is, the ideal single-carrier energy
level of Eq. (\ref{eq14}). Using these populations in the
nondegenerate limit [cf. Eq. (\ref{eq33})], the luminescence
spectrum of Eq. (\ref{eq13}) is given by

\begin{eqnarray}
I(\omega |t) &\propto& \left[ 1+ (\alpha -1) \frac{m_x}{m^*_e}
\mathcal{B}^{e}_{\alpha} (t) \hbar \Omega
\right]^{\frac{\alpha}{1-\alpha }} \left[ 1+ (\alpha -1)
\frac{m_x}{m^*_h} \mathcal{B}^{h}_{\alpha} (t) \hbar \Omega
\right]^{\frac{\alpha}{1-\alpha}} \nonumber \\ &=& \left[ 1+
(\alpha -1) \beta_{eff_\alpha}(t) \hbar \Omega + (\alpha -1)^2
\mathcal{B}^{e}_{\alpha} (t) \mathcal{B}^{h}_{\alpha} (t)
\frac{m^2_x}{m^*_e m^*_h} (\hbar \Omega)^2
\right]^{\frac{\alpha}{1-\alpha}}, \label{eq44}
\end{eqnarray}

\noindent where $\beta_{eff_\alpha}= \frac{m_h^*}{M}
\mathcal{B}^e_\alpha (t) + \frac{m_e^*}{M} \mathcal{B}^h_\alpha
(t)$, $m^{-1}_x = [m^*_e]^{-1}+[m^*_h]^{-1}$, and $M=m^*_e+m^*_h$.
The second contribution in the last term in Eq. (\ref{eq44}) is
much smaller than the first, as verified {\it a posteriori}, and
then the spectrum is approximately described by

\begin{equation}
I(\omega |t) \propto \left[ 1+ (\alpha -1) \beta_{eff_\alpha} (t)
( \hbar \omega - E_G ) \right]^{\frac{\alpha}{1-\alpha }}.
\label{eq45}
\end{equation}

\noindent The experiments reported in Ref. \cite{laureto} are time
integrated, that is, the spectrum is given by

\begin{equation}
\mathcal{I}  (\omega) = \frac{1}{\Delta t} \int_{t}^{t+\Delta t}
dt'I(\omega|t') \, , \label{eq46}
\end{equation}

\noindent where $\Delta t$ is the resolution time of the
spectrometer. Using Eq. (\ref{eq45}) in Eq. (\ref{eq46}), and in
the spirit of the mean-value theorem of calculus we write

\begin{equation}
I(\omega) \propto \left[ 1+ (\alpha -1) \bar{\beta}_{eff_\alpha} (
\hbar \omega - E_G ) \right]^{\frac{\alpha}{1-\alpha }},
\label{eq47}
\end{equation}

\noindent introducing the mean value $\bar{\beta}_{eff_\alpha}$
(as an open parameter), which we rewrite as
$[\bar{\beta}_{eff_\alpha}]^{-1}  = k_B \Theta_\alpha$, defining
an average, over the resolution time $\Delta t$, effective
temperature of the nonequilibrium carriers, that is, a measure of
their average kinetic energy (see Ref. \cite{nettleton}).

\begin{figure}
\includegraphics[angle=0,width=12cm]{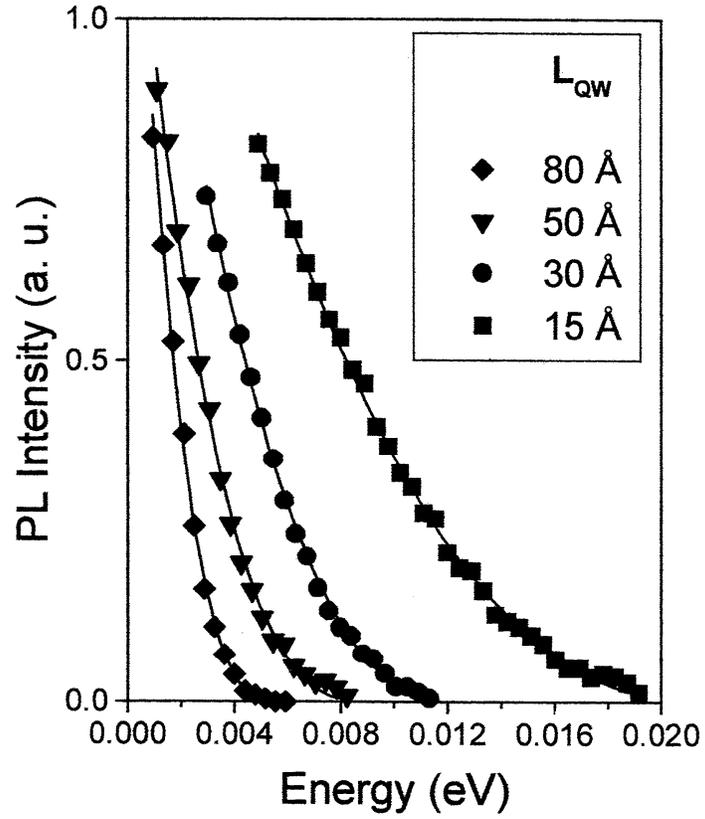}
\caption{The high-energy side of the photoluminescence spectra for
the quantum-well widths indicated in the upper right inset. The
continuous lines are the best fittings to experimental curves
using Eq. (\ref{eq45}).} \label{fig3}
\end{figure}

\begin{figure}
\includegraphics[angle=0,width=12cm]{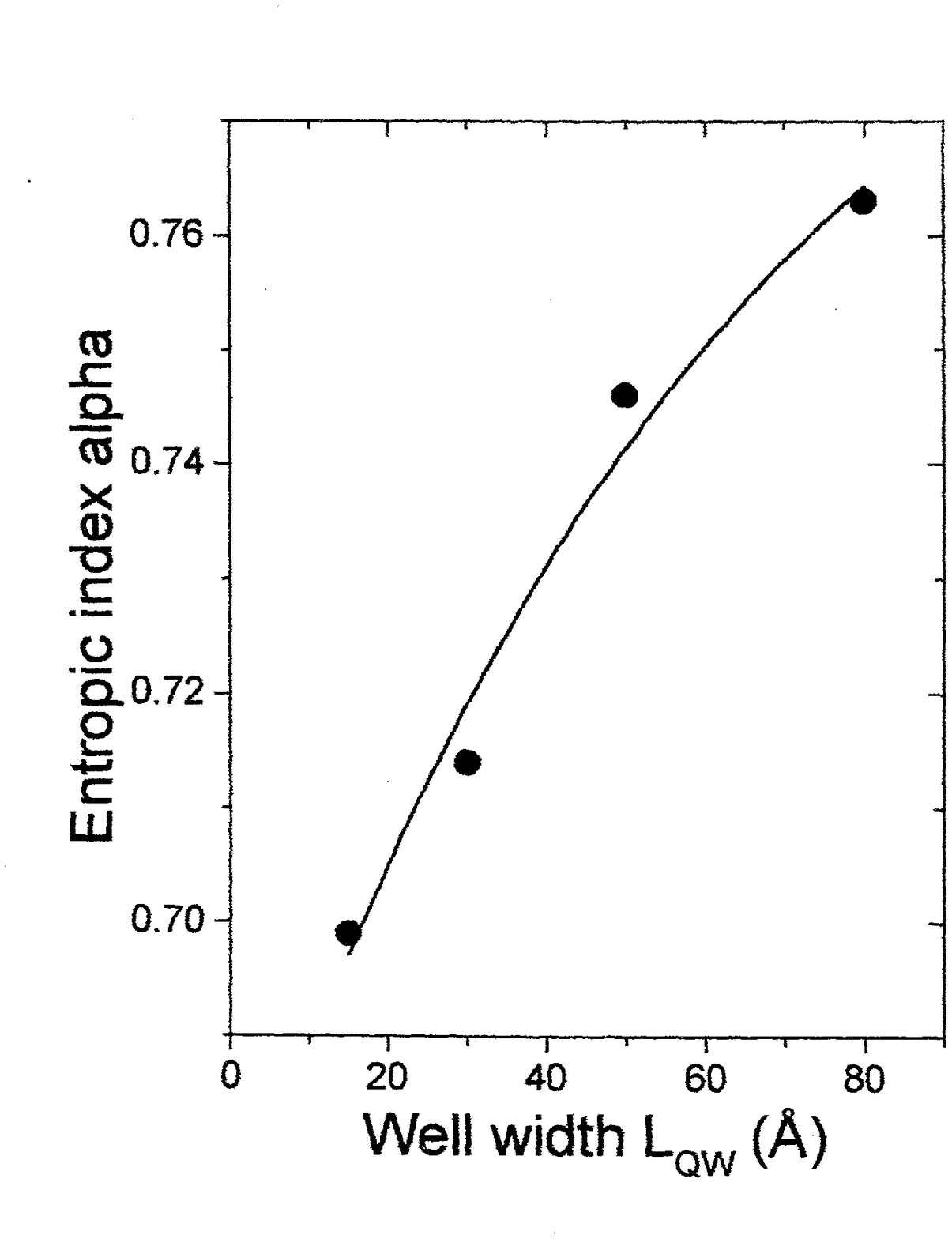}
\caption{Dependence of the infoentropic index $\alpha$ with the
quantum-well width following the approximate empirical law of Eq.
(\ref{eq48}).} \label{fig4}
\end{figure}

In Fig. \ref{fig3} is shown the fitting of the experimental data
with the theoretical curve as obtained from Eq. (\ref{eq47}). It
contains the results referring to four samples having different
values of the quantum well width. The information-entropic index
$\alpha$ depends, as expected, on the dimensions of the system: as
the width of the quantum well increases the values of $\alpha$
keep increasing and tending to 1 (see Fig. \ref{fig4}). This is a
clear consequence that the fractal-like granulation of the
boundary surface becomes less and less relevant for influencing
the outcome of the phenomenon, as the width of the quantum well
falls outside the nanometer scale, and is approached the situation
of a normal bulk sample. On the other hand the kinetic temperature
of the carriers is smaller with increasing quantum well width, as
also expected once the relaxation processes, mainly as a result of
the interaction with the phonon system, become more effective and
the cooling down of the hot carriers proceeds more rapidly.

Moreover, it can be empirically derive what we term as a law of
{\it `` path to sufficiency''}, namely,

\begin{equation}
\alpha (L) \simeq \frac{L+L_1}{L+L_2},
\label{eq48}
\end{equation}

\noindent where, by best fitting, $L_1 \simeq 139 \pm 17$, $L_2
\simeq 204 \pm 24$, all values given in \AA ngstrom. We do have
here that as $L$ largely increases, the entropic index tends to 1,
when one recovers the expressions for the populations in the
conventional situation, but as $L$ decreases $\alpha$ tends to a
finite value $L_1/L_2$, in this case $\simeq 0.7 \pm 0.06$. This
indicates that the insufficiency of description when using Eq.
(\ref{eq14}) in the calculations (the ideal energy levels) becomes
less and less relevant as the size of the system increases as
already commented.

This illustration of the theory clearly evidences the already
stated fact that the infoentropic index $\alpha$ is not a
universal one for a given system, but it depends on the knowledge
of the correct dynamics (the region of energy-momentum space that
is involved in the experiment being analyzed), the geometry and
size including the characteristics and influence of the boundary
conditions (e.g. the fractality in the present case), the
macroscopic (thermodynamic) state of the system in equilibrium or
nonequilibrium conditions, and the experimental protocol.

The case we presented consisted of experiments in time-integrated
optical spectroscopy. The phenomenon of ``anomalous'' luminescence
in nanometric quantum wells in semiconductor heterostructures is
also present in the case of time-resolved experiments (nanosecond
time resolution where the infoentropic index and the
(nonequilibrium) kinetic temperature change in time accompanying
the irreversible evolution of the system \cite{aurea;brasil}). The
use of USM in a completely analogous way as done above (note that
Eq. (\ref{eq44}) is valid for a time-resolved situation) allows to
determine the evolution in time of the kinetic temperature
$\Theta_\alpha (t)$, and the {\it infoentropic index} $\alpha
(t)$, {\it which is then changing in time} as it accompanies the
evolution of the irreversible processes in the nonequilibrium
thermodynamic state of the carriers.

This case involves, as noticed, the difficulty of {\it
insufficiency} in the characterization of the energy levels of the
carriers in the quantum well, i.e. it is of a microscopic
mechanical character. As shown before, use of the escort
probability introduces {\it ad hoc} information trying to account
for missing relevant correlations in the problem, and Renyi
(heterotypical) statistical operator produces modifications in the
carriers' distributions trying to account for insufficiency of
knowledge about the proper quantum mechanical levels.

\subsection{"Anomalous" Diffusion}

On the other side, insufficiency of description at a macroscopic
level is present in questions involving hydrodynamics. So called
``anomalous'' situations are also presented here, e.g.
``anomalous'' diffusion (non-Fickian diffusion)
\cite{crank,aurea;jou}. The question now is which is the source of
the {\it lack of sufficiency} in the well established
Boltzmann-Gibbs formalism, which forces us to resort to the
unconventional approaches. The answer resides in that is being
used a quite incomplete hydrodynamic approach in situations which
require a more detailed treatment. In its more general approach,
the description of hydrodynamical (including rheological) motion
should consist of an extended description, which can be
referred-to as Non-Linear Higher-Order Hydrodynamics with
fluctuations \cite{aurea;nonlin,jou}. Ignoring fluctuations
(relevant for example in turbulent motion) one needs to introduce
the densities of energy, $h(\mathbf{r},t)$, and particles,
$n(\mathbf{r},t)$, and their fluxes of all order, namely
$\mathbf{I}_{p} (\mathbf{r},t)$, for the first (vectorial) fluxes,
${I}_{p}^{[r]} (\mathbf{r},t)$ for the higher-order ($r$-rank
tensor) fluxes, $r \ge 2$, and $p=h$ or $n$ for energy and
particle (material) motion respectively. The motion is then
determined by a complicated set of equations of evolutions of the
type \cite{luzzi1,luzzi51,jou,dedeu}

\begin{equation}
\frac{\partial}{\partial t} I_p^{[r]} (\mathbf{r},t) + \nabla
\cdot I_p^{[r+1]} (\mathbf{r},t) = \mathcal{J}_p^{[r]}
(\mathbf{r},t) \, , \label{eq49}
\end{equation}

\noindent where $r=0$ for the density, $r=1$ for the first
(vectorial) flux, or current, and $r \ge 2$ for the all
higher-order fluxes, $\mathcal{J}_n^{[r]}$ are collision
integrals, and $\nabla \cdot$ is the operator indicating to take
the divergence of the tensor. Solving this set of coupled
equations of evolution is a formidable, almost unmanageable, task.
As a rule one uses, depending on each experimental situation, a
truncation on the set of equations (i.e. they are considered from
$r=0$ up to a certain value, say $n$, of the order $r$)
\cite{RVL00b}. Hence, if one is restricted to introduce a low
order truncation it is faced a failure of the {\it criterion of
sufficiency} when using the conventional, and universal, approach.
Consequently, a way to circumvent the difficulty is, as shown
before, to make calculations in term of unconventional statistical
mechanics, e.g. resorting to Renyi's approach.

The necessity to go over higher orders comes, for example, because
of the presence of a fractal-like structure in the system. We have
dealt in detail the question of polymeric solutions
\cite{aurea;jou}, where the question is presented and discussed in
detail being shown how there follows in USM ``anomalous''
Maxwell-Cattaneo and diffusion equations. In this case,
macromolecules under flow, one faces the difficulty in the
description resulting from the self-similarity in a type of
average fractal structure that this shows (what we have called
``Jackson Pollock Effect'', in view of the analogy with his
paintings with the dripping method, showing to fractal structure
\cite{rptaylor}). Without entering into details, given elsewhere
\cite{aurea;jou}, the ``anomalous'' diffusion equation, derived in
Renyi statistics, has the form

\begin{equation}
\frac{\partial}{\partial t} n (\mathbf{r},t) - D_\alpha
(\mathbf{r},t) \nabla^2 n^{\gamma_\alpha} (\mathbf{r},t) = -
\tau_\alpha (\mathbf{r},t) \nabla \cdot \nabla \cdot (n
(\mathbf{r},t) [\mathbf{v} (\mathbf{r},t) \mathbf{v}
(\mathbf{r},t)])+...\, , \label{eq50}
\end{equation}

\noindent where $D_\alpha (\mathbf{r},t)$ is a transport
coefficient whose dependence on position and time is a consequence
of its dependence on the nonequilibrium thermodynamic state of the
system, and so is the case of the momentum relaxation time
$\tau_\alpha (\mathbf{r},t)$; moreover, the first term on the
right -- depending on the velocity field -- is the double
divergence of the so-called convective pressure tensor
$([\mathbf{v} \mathbf{v}]$ stands for tensorial product of twice
the velocity vector rendering a rank-2 tensor), and we have
omitted to write down additional terms involving gradients of the
transport coefficient and the momentum relaxation time. Finally,
the power $\gamma_\alpha$ is

\begin{equation}
\gamma_\alpha = (5-3\alpha) / (3 - \alpha)\, ,
\label{eq51}
\end{equation}

\noindent with $\alpha$ being, we recall, the infoentropic index
in Renyi statistics, which is limited to the interval $1 \le
\alpha < 5/3$.

``Anomalous'' diffusion is also called-for the explanations of the
results of measurements in the case of experiments of cyclic
voltammetry in microbatteries with fractal-like structured
electrodes \cite{bard}: As a consequence of the nowadays large
interest associated to the development of microbatteries, the
study of growth, annealing, and surface morphology of thin-films
depositions used in cathodes, has acquired particular relevance
\cite{julien}. These kind of systems are characterized by
microroughnessed surfaces in a geometrically constrained region
(nanometer scale), and then fractal characteristics can be
expected to greatly influence the physical properties
\cite{family}. In cyclic voltammetry the resulting current between
electrodes is determined by the charges arriving at the fractal
electrode in a process of hydrodynamic motion. Characterizing such
motion as a Fickian diffusion fails to explain the experimental
data, what can be done using a postulated law of  ``anomalous''
diffusion \cite{bard}. The use of the latter is, as noticed
before, a fitting procedure in  a situation that requires a
higher-order thermo-hydrodynamics (motion of the fluid of charges
in the nano- and subnanometric spaces around the surface of the
fractal-like structured electrode). The use of a zeroth-order
thermo-hydrodynamics -- the one leading to Fickian diffusion
equations in a BG thermo-statistics -- fails to satisfy Fisher's
criterion of sufficiency, and if one persists in using a
zeroth-order hydrodynamics, it needs be handled in an
unconventional approach. Such study has been performed resorting
to Renyi statistics with details given in Refs.
\cite{aurea6247,aurea;gorenstein}, leading to Eq. (\ref{eq50}).
Here we only noticed that, keeping fixed all physical
characteristics in the experiment, if the morphology of the
fractal electrode surface is modified, it can be established a
relation between  Renyi infoentropic index and the average fractal
dimension $d_f$ in the form

\begin{equation}
\alpha = (4 d_f -7) / (2 d_f -3)\, .
\label{eq52}
\end{equation}

\noindent We recall that values of $\alpha$ are permitted in the
interval $1 \le \alpha < 5/3$ (otherwise the theory gives rise to
singularities), now a clear physical restriction once then $2 \le
d_f <3$ as it should.

\subsection{Ideal Gas in Finite Box}

Let us now consider a simple but quite clarifying case, namely an
ideal gas in a finite box (details in Ref. \cite{aurea6247}).
According to the exact result in R. K. Pathria textbook,
\cite{pathriabook} and see also \cite{pathriapaper}, in the case
of a finite but large box there follows for the energy per particle
in the semiclassical limit the expression

\begin{equation}
\frac{E}{N} \simeq \frac{3}{2} k_B T \left[ 1 + \frac{1}{12}
\frac{A \lambda_T}{V} \right], \label{eq53}
\end{equation}

\noindent where $V$ is the volume and $A$ the area of the box, and
$\lambda_T = \hbar / \sqrt{m k_B T}$ is the mean thermal de
Broglie wavelength of the particles in the grand-canonical
ensemble at temperature $T$. Evidently, in the thermodynamic limit
it is recovered the result of energy equipartition. On the other
hand, calculating in the thermodynamic limit but introducing the
equivalent of the grand-canonical distribution in Renyi approach,
in order to account for the insufficiency of description, we
obtain that

\begin{equation}
\frac{E}{N} \simeq \frac{3}{2} F_{h \alpha}^{-1} \left[ 1 +
\frac{1}{4} (\alpha -1)\right], \label{eq53b}
\end{equation}

\noindent where $F_{h \alpha}$ is the Lagrange multiplier
associated to the energy and then, equating  Eqs. (\ref{eq53}) and
(\ref{eq53b}) one finds that

\begin{equation}
\alpha \simeq 1 + 4 \left[  k_B T F_{h \alpha} \left(  1 +
\frac{1}{12} \frac{A \lambda_T}{V}\right)-1 \right]. \label{eq54}
\end{equation}

\noindent In the thermodynamic limit $A/V$ goes to zero and
$F^{-1}_{h \alpha}$ goes to $k_B T$ and then $\alpha$ goes to 1 as
it should. Moreover, from Eq. (\ref{eq54}) we can clearly see that
the infoentropic index $\alpha$ depends on the system dynamics,
its geometry and size, and the thermodynamic state.

\subsection{Black Body Radiation in Insufficient Description}

Finally, we briefly mention the case of black-body radiation in
insufficient description. We consider the gas of photons in the
presence of an uniform flux of energy (generated, for example, as
a result of the presence of different temperatures at both ends of
the container). We look for the calculation of the energy what is
done, on the one side, using the conventional approach in a
description that includes the energy and the energy flux as basic
variables (that is, in a first-order thermo-hydrodynamics; see
discussion on higher-order thermo-hydrodynamics earlier in this
Section) and, on the other side, using  an incomplete description
including only the energy (zeroth-order thermo-hydrodynamics) but
dealt with using Renyi statistics trying to compensate for the
insufficiency of description. Details are given in Ref.
\cite{aurea6247}, and here we only notice that equating the values
of the energy in both descriptions, it follows that the
infoentropic index is given by

\begin{equation}
\alpha = 1 - (I/I_0)^2 ,
\label{eq55}
\end{equation}

\noindent where $I$ is the energy flux and $I_0^2 = (8/15 V^2) a^2
c^2 T^8$ with $a$ being Stefan-Boltzmann constant. It can be
noticed that $I_0$ is a kind of a flux of energy composed of the
energy density of the radiation, $a T^4/V$, traversing at the
speed of light, while $I$ is roughly the density of energy
traversing at a speed determined by the gradient of temperature,
and then $I/I_0$ is very small, and then the infoentropic index is
practically 1 meaning that the insufficiency in the
characterization can be ignored.

\section{Concluding Remarks}

However the enormous success and large application of
Shannon-Jaynes approach to derive the probability distributions of
the ensemble formalism in the Laplace-Maxwell-Boltzmann-Gibbs
statistical foundations of physics, as it has been noticed, some
cases are difficult to be properly handled within such
formulation, as a result of existing some kind of fuzziness in
data or information, that is, the presence of conditions of
insufficiency in the characterization of the (microscopic,
macroscopic or mesoscopic) state of the system. Such difficulty
with the proper characterization  of the system in the problem in
hands can be somehow compensated, as shown, with the introduction
of peculiar parameter-dependent alternative
informational-entropies (see Table I), allowing for the
construction of a vast number of auxiliary heterotypical
statistics. {\it In that way one recovers the ability to make
improved predictions on the properties of the system, and be
capable to get a picture of the physical processes involved}
allowing, as for example illustrated in Section III, to evaluate
the influence of peculiar characteristics of the sample (like,
e.g., its fractal structure) on its physico-chemical behavior.

Returning to the question of the possible failure to satisfy
Fisher's criteria of efficiency and sufficiency, in the case of
particularly dealing with systems with some kind of fractal-like
structure the use of Shannon-Jaynes infoentropy would require to
introduce as information the highly correlated conditions that are
in that case present. Two examples in condensed matter physics
have been described in the previous Section in which fractality
enters via the non-smooth topography of the boundary surfaces
which have large influence on phenomena occurring in constrained
geometries (nanometer scales in the active region of the sample).
Hence the most general and complete Boltzmann-Gibbs formalism in
Shannon-Jaynes approach becomes hampered out and is difficult to
handle, and then, as shown, use of other types of
informational-entropies leads to the derivation of heterotypical
probability distributions on the basis of the constrained
maximization of unconventional informational-statistical entropies
(quantity of uncertainty of information), to be accompanied, as
noticed in the main text, with the use of the so-called escort
probabilities, allowing for an analysis and understanding of the
observed physical behavior of the system.

Summarizing, {\it Unconventional Statistical Mechanics consists of
two steps: 1. The choice of a deemed appropriate structural
informational-entropy for generating the heterotypical statistical
operator, and 2. The use of a escort probability in terms of the
heterotypical distribution of item 1.}

As shown in the main text, and illustrated  in Section III, the
{\it escort probability} introduces corrections to an inefficient
description by including correlations and higher-order variances
of the observables involved. On the other hand, the {\it
heterotypical distribution} introduces corrections to the
insufficient description by modifying the statistical weight of
the dynamical states of the conventional approach involved in the
situation under consideration. Moreover, we have considered a
particular case, namely the statistics as derived from the use of
Renyi informational entropy, centering the attention on the
derivation of an Unconventional Statistical Mechanics appropriate
for dealing with far-removed-from-equilibrium systems. Moreover,
we have reported the calculation, in such conditions, of the
distribution functions of single fermions and bosons, the
counterpart in these unconventional statistics of the usual
Fermi-Dirac and Bose-Einstein distributions: These distributions
are illustrated in Figs. \ref{fig1} and \ref{fig2}.

In conclusion, we may say that USM appears as a valuable approach,
in which the introduction of
informational-entropic-indexes-dependent informational-entropies
leads to a particularly convenient and sophisticated method for
studying certain classes of physical systems for which the
criteria of efficiency and sufficiency in its characterization
cannot be properly satisfied.

Finally, it is relevant to emphasize again the fact that the
infoentropic index(es) is(are) dependent on the dynamics involved,
the system's geometry and dimensions, boundary conditions, its
macroscopic-thermodynamic state (in equilibrium, or out of it when
becomes a function of time), and the experimental protocol.

Also, we recall the fundamental point, stressed in the
Introduction, that these informational entropies (informational
measures) are not at all to be interpreted as the physico-thermal
entropy of the system.

\section{Acknowledgments}

We acknowledge financial support to our Group provided in
different opportunities by the S\~ao Paulo State Research
Foundation (FAPESP), the Brazilian National Research Council
(CNPq), the Ministry of Planning (Finep), the Ministry of
Education (CAPES), Unicamp Foundation (FAEP), IBM Brasil, and the
John Simon Guggenheim Memorial Foundation (New York, USA).

\newpage

\end{document}